\documentclass[11pt]{article}
\usepackage{amsmath,amssymb,amsfonts,amsthm,epsfig}
\usepackage[usenames,dvipsnames]{xcolor}
\usepackage{bm,xspace}
\usepackage{tcolorbox}
\usepackage{cancel}
\usepackage[margin=0.8in]{geometry}
\usepackage{liyang}
\usepackage{framed}
\usepackage{verbatim}
\usepackage{enumitem}
\usepackage{array}
\usepackage{multirow}
\usepackage{afterpage}
\usepackage{mathrsfs}
\usepackage{pifont} 
\usepackage{chngpage}
\usepackage[normalem]{ulem}
\usepackage{boxedminipage}
\usepackage{caption}
\usepackage{subcaption}
\usepackage{bold-extra}
\usepackage{forest}
\usepackage{etoolbox}
\usepackage{thm-restate}
\usepackage{booktabs,array}
\usepackage{microtype}
\usepackage{subcaption}
\usepackage{algorithm}
\usepackage{algorithmicx}
\usepackage{algpseudocode}

\usepackage{tikz}
\usetikzlibrary{calc,through,backgrounds,decorations.pathreplacing, calligraphy,arrows.meta}
\usetikzlibrary{positioning,chains,fit,shapes}
\usetikzlibrary{patterns}

\usepackage{tikz-cd}
\usepackage{twemojis}

\usepackage{pgfplots}
\pgfplotsset{width=8cm,compat=newest}

\def\colorful{1}

\ifnum\colorful=1

\fi
\ifnum\colorful=0

\fi

\newcommand{\NP}{\mathsf{NP}}

\renewcommand{\P}{\mathsf{P}}

\newcommand{\NSPACE}{\mathsf{NSPACE}}

\DeclarePairedDelimiterXPP{\DTIME}[1]{\mathsf{DTIME}}{[}{]}{}{#1}
\DeclarePairedDelimiterXPP{\NTIME}[1]{\mathsf{NTIME}}{[}{]}{}{#1}

\newcommand{\timefn}{t}
\newcommand{\blocksize}{b}
\Crefname{theorem}{Theorem}{Theorems}
\Crefname{thmenumi}{Theorem}{Theorems}
\AtBeginEnvironment{theorem}{%
    \crefalias{enumi}{thmenumi}%
    \setlist[enumerate,1]{
        label={\textit{(\roman*)}},
        ref={\thetheorem(\roman*)}
    }%
}

\colorlet{headtriangle}{gray!50!red}
\newcommand{\circuit}[5]{%
    \coordinate (#1bl) at ({#2-#3},1.75);
    \coordinate (#1br) at ({#2+#3},1.75);
    \coordinate (#1ap) at (#2,3.0);
    \draw[thick, fill=gray!10, line join=round] (#1bl) -- (#1br) -- (#1ap) -- cycle;
    \node[font=\small] at (#2,2.17) {$C_{#4}$};
    \node[font=\scriptsize, above] at ($(#1ap)+(0,0.25)$) {$|S_{#4}|=#5$};}
\newcommand{\wires}[3]{% name, count, list of variables
    \foreach \v [count=\k] in {#3}
      \draw[wire] ($(#1bl)!{(\k-0.5)/#2}!(#1br)$) -- (x\v.north);}

\newcommand{\prev}{\mathrm{prev}}

\begin{document}

\title{
 Simulating Nondeterministic Turing Machines \\ in Subexponential Deterministic Time}

\author{ 
 Carmen Strassle  \\
 \hspace{-5pt} { {\sl Stanford University}} \vspace{15pt}
 \and 
Bernardo Subercaseaux   \\
\hspace{-10pt} {{\sl Carnegie Mellon University}}
}

\date{\small{\today}}

 \maketitle

\begin{abstract}
We show that every multitape nondeterministic Turing machine can be simulated deterministically in subexponential time. Specifically, we show that the recent algorithms for~$3$-$\mathsf{SAT}$ presented by OpenAI and Anthropic~imply $\NTIME{\timefn}\subseteq\DTIME*{2^{O(\timefn \log \log \timefn/\sqrt{\log \timefn})}}$, and that a stronger consequence of their proof yields \[\NTIME{\timefn} \subseteq\DTIME*{2^{O(\timefn /\sqrt{\log \timefn})}}.
\]
More generally, we show that any subexponential time algorithm for solving circuit-CSPs transfers to a deterministic simulation of $\mathsf{NTIME}$ with the \emph{same} savings. Our results follow from new generalizations of Cook--Levin, which may be of independent interest. 
\end{abstract} 

 \setcounter{page}{1}

\section{Introduction}
The exponential time hypothesis (ETH) of Impagliazzo and Paturi~\cite{IP01} has been recently refuted:
\begin{fact}[\cite{openai26, Claude2026ETH}]\label{fact:ksat}
    $3$-$\mathsf{SAT}$~over $n$ variables and $O(n)$ clauses can be solved in time $2^{O\big(n/\sqrt{\log n}\big)}$.
\end{fact}

It's natural to ask whether this result implies a more general speed-up for other problems that can be solved efficiently with nondeterminism. However, the most direct attempt to show $\NTIME{\timefn}\subseteq\DTIME*{2^{o(\timefn)}}$ based on~\Cref{fact:ksat} fails: the most efficient known version of Cook--Levin reduces acceptance of a nondeterministic Turing machine running in time $\timefn$ to the satisfiability of a 3-CNF formula over $O(\timefn \log \timefn)$ variables and clauses~\cite{cook1988short}. Plugging this into~\Cref{fact:ksat} yields $\NTIME{\timefn}\subseteq\DTIME{2^{O(\timefn\sqrt{\log \timefn})}},$ which is worse than the trivial upper bound of $\DTIME{2^{O(\timefn)}}$. So this proof strategy would require improving upon the parameters of quasilinear Cook--Levin, which is a long-standing open problem. 

\begin{figure}[t]
\centering

\begin{tikzpicture}[
    scale = 1,
    var/.style={circle, draw, thick, minimum size=0.62cm, inner sep=0pt, font=\small, fill=white},
    con/.style={draw, thick, rounded corners=2pt, minimum width=1.25cm, minimum height=0.7cm, font=\small, fill=gray!40},
    wire/.style={Latex-, thick, color=black!90},
  ]
  \foreach \i in {1,...,8} \node[var] (x\i) at ({1.1*\i},0) {$x_{\i}$};

  \circuit{C1}{2.2}{0.8}{1}{4}
  \circuit{C2}{4.4}{0.62}{2}{3}
  \circuit{C3}{6.6}{0.95}{3}{5}
  \circuit{C4}{8.6}{0.45}{4}{2}
  \wires{C1}{4}{1,2,3,4}
  \wires{C2}{3}{3,5,6}
  \wires{C3}{5}{4,5,6,7,8}
  \wires{C4}{2}{7,8}

  \node[font=\small, anchor=east] at (0.35,2.3) {constraints};
  \node[font=\small, anchor=east] at (0.35,0) {variables};
\end{tikzpicture}
\caption{A circuit-CSP over $n=8$ variables and $m=4$ constraints of arity $5$ and density $14/8$.}
\label{fig:csp}
\end{figure}
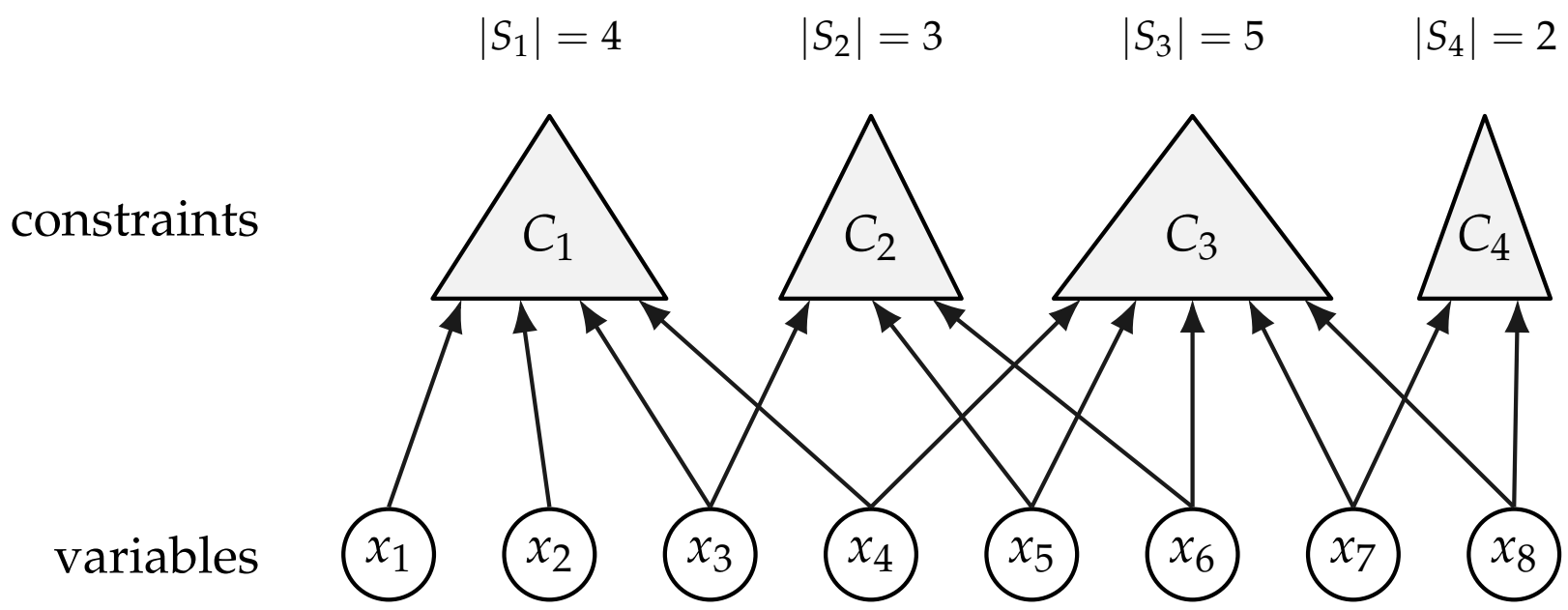

In this work, we show that the recent ETH refutation does indeed apply to the general simulations of nondeterministic computation---even with the same savings: 
 \begin{tcolorbox}[colback = white,arc=1mm, boxrule=0.25mm]
\begin{theorem}[Subexponential Simulation of NTMs]\label{thm:lossless} 
    For every growth function $\timefn\colon \mathbb{N} \to \mathbb{N}$ such that $\timefn(n) \geq n$, we have \[\NTIME{\timefn}\subseteq\DTIME*{2^{O\left({\timefn}/{\sqrt{\log \timefn}}\right)}}.\]
\end{theorem}  
\end{tcolorbox}
\medskip
We obtain~\Cref{thm:lossless} as a corollary of a more general ``transfer'' theorem showing how 
further improvements in algorithms for SAT yield further improvements in the $\mathsf{NTIME}$ vs.~$\mathsf{DTIME}$ setting. To state this transfer theorem, we first define constraint-satisfaction problems (CSPs) where each constraint is represented by a circuit (see~\Cref{fig:csp}).

\begin{definition}[Circuit-CSP]
A \emph{circuit-CSP} $\Phi$ consists of Boolean variables ${x} := x_1,\dots,x_n$ together with constraints $(S_1,C_1),\dots,(S_m,C_m)$, where each $S_j$ is a tuple of distinct variables from ${x}$ and each $C_j$ is a Boolean circuit on $|S_j|$ inputs. An assignment to ${x}$ satisfies $\Phi$ if $C_j(S_j)=1$ for all $j$. The \emph{arity} of $\Phi$ is $\max_j|S_j|$, and its \emph{density} is $\frac1n\sum_j|S_j|$.
\end{definition}

For instance, if all $C_j$ are disjunctions of $k$ literals (variables or negations thereof), then this definition corresponds to a $k$-CNF. Generally, larger arity and density make it more difficult to decide if a circuit-CSP is satisfiable. 

\noindent Stated in terms of circuit-CSPs, our  transfer theorem is as follows:

\begin{tcolorbox}[colback = white,arc=1mm, boxrule=0.25mm]
\begin{theorem}[Circuit-CSP solvers imply NTM simulators]\label{thm:lossless-general}
Let $t(n),\, k(n),$ and $s(n)$ be growth functions such that $\timefn(n)\geq n$ and  $2\leq s(n) \leq n/\log n$. Suppose that circuit-CSP over $n$ variables with arity at most $k(n)$ and density at most $2$ can be solved in time $2^{O(n/s(n))}\poly(|\Phi|)$, then 
    \[
\NTIME{\timefn}\subseteq\DTIME*{2^{O(\timefn/k(\timefn)) } \cdot 2^{O(\timefn/s(\timefn)) }}.
\]
\end{theorem}
\end{tcolorbox}
Notice that as long as $k(n) \geq \Omega(s(n))$, the above theorem \emph{losslessly} transfers any improvements in a circuit-CSP solver to an improved upper bound for $\mathsf{NTIME}$.
In order to recover~\Cref{thm:lossless} from~\Cref{thm:lossless-general}, we need a circuit-CSP version of~\Cref{fact:ksat}. We show that~\cite[Theorem D.1]{openai26} and the analogous~\cite[Lemma 8 and Theorem A]{Claude2026ETH} imply the following:
\begin{fact}[\cite{openai26,Claude2026ETH}]\label{fact:csp}
    Circuit-CSP over $n$ variables with arity at most $n^{9/10}$ and constant density can be solved in time $2^{O(n/\sqrt{\log n})}\poly(|\Phi|)$.
\end{fact}

\noindent Combining this fact with~\Cref{thm:lossless-general} directly recovers~\Cref{thm:lossless}. 

It is worth noting that~\Cref{fact:csp} is stronger than~\Cref{fact:ksat}. Naturally, one would hope for an improvement to the state of $\mathsf{NTIME}$ vs.~$\mathsf{DTIME}$ even if we are only able to improve the algorithm for $3$-SAT and not circuit-CSP. We therefore provide a variant of~\Cref{thm:lossless-general} that exchanges an easier-to-satisfy assumption for weaker parameters:

\begin{tcolorbox}[colback = white,arc=1mm, boxrule=0.25mm]
\begin{theorem}[$3$-SAT solvers imply NTM simulators]\label{thm:lossy-general}
    Let $\timefn(n)\geq n$ and $2\leq s(n) \leq n/\log n$ be growth functions. Suppose that 3-SAT over $n$ variables and $O(n)$ clauses can be solved in time $2^{O(n/s(n))}$, then 
    \[
\NTIME{\timefn}\subseteq\DTIME*{2^{O(\timefn\log s(\timefn) /s(\timefn)) }}.
\]
\end{theorem}
\end{tcolorbox}
This conclusion is weaker than~\Cref{thm:lossless-general} since it has an extra factor of $\log s(\timefn)$ in the exponent. Setting $s(n)=\sqrt{\log(n)}$ and applying~\Cref{fact:ksat} gives the following corollary, which is a weaker version of~\Cref{thm:lossless}: $\NTIME{\timefn}\subseteq\DTIME{2^{O(\timefn\log \log \timefn /\sqrt{\log \timefn}) }}.$

\begin{remark}
\cite{openai26}'s Theorem D.1 is stronger than~\Cref{fact:ksat} or even~\Cref{fact:csp}, as it solves the more difficult problem of Boolean tensor contraction (see~\Cref{def:tensor}). Our goal when formulating~\Cref{thm:lossless-general} in terms of circuit-CSP was to
choose a ``minimal'' problem for which algorithmic improvements would translate losslessly to~$\mathsf{NTIME}$.

Relatedly, \cite{openai26, Claude2026ETH} further prove that their algorithms count the number of satisfying assignments (e.g.,~they solve \#circuit-CSP). With this strengthened assumption and some additional care in the argument, all of our deterministic algorithms can also produce a count of the number of accepting paths. 
\end{remark}

\subsection{Background and context} 
Prior to this work, very little was known about the relationship between $\mathsf{NTIME}$ and $\mathsf{DTIME}$.
Paul, Pippenger, Szemerédi, and Trotter~\cite{PPST83} showed the modest separation $\NTIME{n}\neq \DTIME{n}$, which was then improved to $\NTIME{n\sqrt{\lg^*(n)}}\neq \DTIME{n\sqrt{\lg^*(n)}}$ by Santhanam~\cite{Santhanam01}. For upper bounds, the question we study here, even less was known. Lipton conjectured that for any $\eps >0,$ 
\[\NTIME{t}\subseteq\DTIME{2^{\eps t}},\]  which has remained an open question and has been asked multiple times in the literature~\cite{Lipton09, Mor81,KLRS12,Doty10}. Some progress was made by~\cite{KLRS12}, who proved a $\DTIME{2^{c t}}$ upper bound with an improvement in the constant $c$. \Cref{thm:lossless} settles Lipton's conjecture in the affirmative by a large margin. 

It's worth noting that there exists an oracle relative to which Lipton's conjecture is \emph{false}~\cite{BGS75,Mor81,Doty10}. Therefore, our proof, by necessity, does not relativize. Furthermore, Bennett and Gill's proof that $\P \neq \NP$ relative to a random oracle~\cite{BG81} can be straightforwardly extended to show that Lipton's conjecture is false even relative to a \emph{random} oracle. Therefore, our result is an example of a natural statement whose truth differs in the random-oracle and unrelativized worlds (cf.~\cite{ChangCGHHRR94}).

\paragraph{Single-tape Turing machines.} For single-tape Turing machines, the situation is much simpler, and we have the following chain of inclusions:
\[
\mathsf{NTIME}_1[t] \subseteq \NSPACE_1\!\left[O(\sqrt{t})\right] \subseteq \DTIME*{2^{O(\sqrt{t})}} \subseteq \mathsf{DTIME}_1\!\left[2^{O(\sqrt{t})}\right],
\]
where the subscript $1$ indicates single-tape machines. The first inclusion was proven by Hopcroft and Ullman~\cite{HU68} (and strictly speaking requires $t(n) \geq n^2$), the second uses the standard configuration graph technique~\cite{Cook71PDA,Savitch70}, and the last one uses the quadratic blow-up for multitape simulations~\cite{HS65}.
Unfortunately,
we don't know how to extend time vs. space results that hold under robust models of computation (e.g., multitape or RAM) to the nondeterministic setting, and thus they do not imply anything about $\mathsf{NTIME}$ vs.~$\mathsf{DTIME}$~\cite{HPV77,Williams25}.

\section{Technical Overview}

Our main technical lemma is a generalized version of Cook--Levin that, instead of mapping a nondeterministic Turing machine to a single 3-CNF, maps it to a \emph{collection} of circuit-CSPs:

\begin{lemma}[Generalized Cook--Levin for CSPs]\label{lem:cl-for-CSP}
    Let $M$ be a nondeterministic multitape Turing machine that runs in time $\timefn(n)$, and let $\blocksize \colon \mathbb{N} \to \mathbb{N}$ be such that $\blocksize(n) \leq \timefn(n)$. Then, there is a deterministic algorithm that takes an input $x$ to $M$ of length $n$ and
    returns circuit-CSPs $\Phi_1,\ldots,\Phi_m$ for $m = 2^{O(\timefn(n)/\blocksize(n))}$ in time $m \cdot \poly(t(n))$,  such that
    \begin{enumerate}
        \item \label{item:csp-correct} $M$ accepts input $x$ if and only if some $\Phi_\ell$ is satisfiable, and 
        \item each $\Phi_{\ell}$ has $O(\timefn(n))$ variables,  arity $O(\blocksize(n))$, density at most 2, and $O(\timefn(n)/\blocksize(n))$ constraint circuits each of size $O(\blocksize(n) \log \blocksize(n))$. 
    \end{enumerate}
\end{lemma}

We then obtain the following corollary, which looks more like the original Cook--Levin because it maps a nondeterministic Turing machine to a collection of 3-CNFs. We need this corollary to prove~\Cref{thm:lossy-general} since it only assumes an improvement for 3-SAT solvers, not the more general circuit-CSP solvers. 

\begin{corollary}[Generalized Cook--Levin]\label{cor:cl}
    Let $M$ be a nondeterministic multitape Turing machine that runs in time $\timefn(n)$, let $x$ be a length $n$ input to $M$, and let $b(n)\leq \timefn(n)$. There is a deterministic algorithm that maps $x$ to 3-CNF formulas $\varphi_1,\ldots, \varphi_m$ that runs in time $m\cdot\poly (\timefn(n))$ for $m=2^{O(\timefn(n)/b(n))}$ such that 
    \begin{enumerate}
        \item $M$ accepts input $x$ if and only if some $\varphi_\ell$ is satisfiable, and
        \item each $\varphi_{\ell}$ has $O(\timefn(n)\log b(n))$ variables and clauses. 
    \end{enumerate}
\end{corollary}

If we set $b=t$ in the above, an inspection of the proof shows that this results in a single 3-CNF formula $\varphi$ over $O(t\log t)$ variables. Additionally, the $\poly(t)$ runtime can be made $ t\cdot\poly (\log t)$ with some additional care in the argument. We do not take this care since, for the parameter regime that we care about, it does not impact the result. Therefore, this corollary recovers the classical quasilinear Cook--Levin and so is indeed a generalization. See~\Cref{tab:cook-levin} for a detailed comparison between our results and quasilinear Cook--Levin. 

\begin{remark}
    These generalizations of Cook--Levin have a similar flavor to the sparsification lemma of Impagliazzo, Paturi, and Zane~\cite{IPZ01}. In fact,
    in the language of~\cite{IPZ01},~\Cref{cor:cl} with $b(n) := \Theta(1/\varepsilon)$ is a \emph{sub-exponential reduction family} (SERF) from the language $L(M)$, parameterized by runtime,  to $\mathsf{3\text{-}CNF}$ parameterized by the number of clauses. 
\end{remark}

\Cref{thm:lossless-general} and \Cref{thm:lossy-general} are immediate from \Cref{lem:cl-for-CSP} and \Cref{cor:cl}. We prove them in~\Cref{sec:lossless-thm-proofs}. 

% \begin{table}[t]
% \centering
% \renewcommand{\arraystretch}{1.7}
% \setlength{\extrarowheight}{1pt}
% \small
% \begin{tabular}{>{\raggedright\arraybackslash}p{2.7cm}|>{\raggedright\arraybackslash}p{2.0cm}|>{\centering\arraybackslash}p{1.45cm}|>{\centering\arraybackslash}p{1.75cm}|>{\raggedright\arraybackslash}p{2.65cm}|>{\centering\arraybackslash}p{2.25cm}}
% \toprule 
%  & \textbf{Object produced} & \textbf{ \#Objects} & \textbf{Variables each} & \textbf{Constraints each} & \textbf{Runtime} \\
% \midrule
% Classical \mbox{Cook--Levin} \cite{cook1988short}
%   & 3-CNF
%   & $1$
%   & $O(t\log t)$
%   & $O(t\log t)$ clauses of width $3$
%   & $\poly(t)$ \\
% \hline
% Generalized Cook--Levin, CSP form (\Cref{lem:cl-for-CSP})
%   & circuit-CSPs $\Phi_\ell$
%   & $2^{O(t/b)}$
%   & $O(t)$
%   & $O(t/b)$ circuits of arity $O(b)$ and size $O(b\log b)$; density $\le 2$
%   & $2^{O(t/b)}\poly(t)$ \\
% \hline
% Generalized Cook--Levin, 3-CNF form (\Cref{cor:cl})
%   & 3-CNFs $\varphi_\ell$
%   & $2^{O(t/b)}$
%   & $O(t\log b)$
%   & $O(t\log b)$ clauses of width $3$
%   & $2^{O(t/b)}\poly(t)$ \\
% \bottomrule
% \end{tabular}
% \caption{Three ways of encoding a nondeterministic time-$t$ Turing machine. In each case, the Turing machine accepts iff some object produced is satisfiable. Taking $b=t$ in the last row recovers the first.}
% \label{tab:cook-levin}
% \end{table}

\begin{table}[t]
\centering
\small
\setlength{\tabcolsep}{3pt}
\begin{tabular*}{\linewidth}{@{\extracolsep{\fill}}
    >{\raggedright\arraybackslash}p{4.05cm} l c c
    >{\raggedright\arraybackslash}p{4.4cm} c @{}}
\toprule
\textbf{Reduction} & \textbf{Output} & \textbf{Count} & \textbf{Variables}
  & \textbf{Constraints} & \textbf{Runtime} \\
\midrule
Classical Cook--Levin\newline {\footnotesize\backrefsetup{disable}\cite{cook1988short}}
  & 3-CNF
  & $1$
  & $O(t\log t)$
  & $O(t\log t)$ clauses of width $3$
  & $t~\poly(\log t)$ \\
\addlinespace
Generalized Cook--Levin\newline {\footnotesize CSP form (\Cref{lem:cl-for-CSP})}
  & circuit-CSPs $\Phi_\ell$
  & $2^{O(t/b)}$
  & $O(t)$
  & $O(t/b)$ circuits of
    arity $O(b)$, size $O(b\log b)$,\newline
    and density $\le 2$
  & $2^{O(t/b)}\poly(t)$ \\
\addlinespace
Generalized Cook--Levin\newline {\footnotesize 3-CNF form (\Cref{cor:cl})}
  & 3-CNFs $\varphi_\ell$
  & $2^{O(t/b)}$
  & $O(t\log b)$
  & $O(t\log b)$ clauses of width $3$
  & $2^{O(t/b)}\poly(t)$ \\
\bottomrule
\end{tabular*}
\captionsetup{font=small,labelfont=bf}
\caption{Three encodings of a nondeterministic time-$t$ Turing machine.
  The machine accepts iff some output is satisfiable. Variable and constraint
  bounds are per output. At $b=t$, the last row recovers the first.}
\label{tab:cook-levin}
\end{table}

\subsection{Proof overview of~\Cref{lem:cl-for-CSP} and \Cref{cor:cl}}\label{subsec:overview}

The typical proof of classical quasilinear Cook--Levin can be described in 2 steps:
\begin{enumerate}
    \item Use a result of~\cite{PippengerF79} to convert the nondeterministic TM into a size $O(t \log t)$ circuit $C$, and then, 
    \item use a result of~\cite{Tseitin68} to convert this circuit $C$ into a 3-CNF $\varphi$ over $O(t\log t)$ variables. 
\end{enumerate}

The $t\log t$ bound on the number of variables results from encoding the entire Turing machine as a single large circuit $C$. Suppose that instead, we were somehow able to break the Turing machine up into size $b$ segments and encode each segment individually. This would result in $O(t/b)$ circuits each of size $O(b\log b)$. Then, when we reach step 2, we would have a 3-CNF over $O(t/b)\cdot O(b\log b)= O(t\log b)$ different variables. This is the entire idea of the proof. It turns out that when we break the Turing machine up into size $b$ segments, there are $2^{O(t/b)}$ ways to do so. This is why we end up with $2^{O(t/b)}$ different 3-CNFs in the statement of~\Cref{cor:cl}, and also why we cannot afford to make $b$ arbitrarily small.

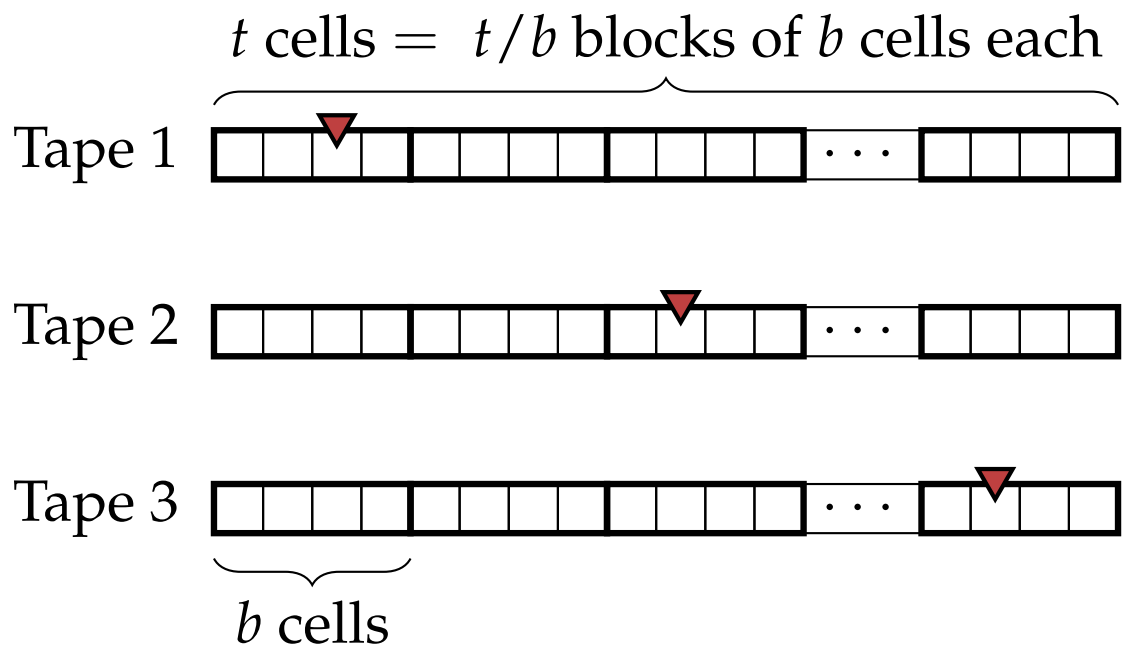
\begin{figure}[ht]
\centering
\begin{tikzpicture}[
    scale=0.65,   % shrinks the drawing; text stays full size
    cell/.style   = {draw, minimum width=0.5cm, minimum height=0.5cm, inner sep=0pt},
    blockbox/.style = {draw, very thick},
    oldhead/.style   = {fill=black, isosceles triangle, isosceles triangle apex angle=60,
                     shape border rotate=90, inner sep=0pt, minimum height=0.18cm},
     head/.style = {draw=black, thick, fill=headtriangle, isosceles triangle, isosceles triangle apex angle=60,
                   shape border rotate=270, inner sep=0pt, minimum height=0.20cm},
    brace/.style  = {decorate, decoration={brace, amplitude=5pt, raise=2pt}},
]

% ---------------- parameters ----------------
\def\cw{0.5}      % cell width
\def\nc{4}        % cells drawn per block (stands for b)
\def\bw{2}        % block width = \nc * \cw
\def\gap{1.2}     % width of the "..." gap between block 3 and block t/b
% x-offsets of the blocks drawn: B_1, B_2, B_3, (gap), B_{t/b}
\def\lastx{6+\gap}

% ---------------- the three tapes ----------------
\foreach \tape/\y in {1/0, 2/-1.8, 3/-3.6} {
  % blocks 1..3
  \foreach \blk in {0,1,2} {
    \foreach \c in {0,...,3}
      \draw ({\blk*\bw + \c*\cw}, \y) rectangle ++(\cw, 0.5);
    \draw[blockbox] ({\blk*\bw}, \y) rectangle ++(\bw, 0.5);
  }
  % gap
  \node at ({3*\bw + \gap/2}, \y+0.25) {$\cdots$};
  \draw (3*\bw, \y) -- ++(\gap, 0);
  \draw (3*\bw, \y+0.5) -- ++(\gap, 0);
  % last block B_{t/b}
  \foreach \c in {0,...,3}
    \draw ({\lastx + \c*\cw}, \y) rectangle ++(\cw, 0.5);
  \draw[blockbox] (\lastx, \y) rectangle ++(\bw, 0.5);

  % tape label (overlay: ignored when centring, so the tapes are centred)
  \node[anchor=east, overlay] at (-0.15, \y+0.25) {Tape \tape};
}

% ---------------- heads ----------------
% head positions (x of cell centres) for each tape
\foreach \y/\hx in {0/1.25, -1.8/4.75, -3.6/7.95} {
  \node[head] at (\hx, \y-0.2 + 0.75) {};
}

\draw[brace, decoration={mirror}] (0, -3.6-0.15) -- node[below=6pt] {$b$ cells} (\bw, -3.6-0.15);

\draw[brace, ] (0, 0.5+0.15)
      -- node[above=6pt] {$t$ cells $=\; t/b$ blocks of $b$ cells each}
      ({\lastx + \bw}, 0.5+0.15);

\end{tikzpicture}
\caption{A 3-tape Turing machine split into $t/b$ blocks of $b$ cells each.
The Turing machine is \emph{block-respecting} if each of its heads stays within
a single block during each length-$b$ epoch.}
\label{fig:block-respecting}
\end{figure}

\paragraph{How to segment an NTM.}  Our technique is similar to that of~\cite[Section 5]{KLRS12}.  We split the runtime of a $t$-time Turing machine $M$ into $\timefn/\blocksize$ \emph{epochs}, each of length $b$. Additionally, we split each of the Turing machine's $d$ tapes into contiguous \emph{blocks}, each consisting of $b$ cells. Then, a result of~\cite{HPV77} (restated in~\Cref{lem:hpv}) allows us to assume without loss of generality that $M$ is \emph{block-respecting}, meaning that it only moves its tape heads from one block to another at the end of an epoch. In other words, for a fixed epoch $i\in [t/b]$ and a fixed tape $j\in [d]$, that tape head will stay within a single block (see~\Cref{fig:block-respecting}). We let $\pi\in \{L,S,R\}^{(t/b-1)\times d}$ be the \emph{pattern} that $M$'s heads take (see~\Cref{fig:pattern}). Specifically, 
\begin{align*}
    \pi_{i,j} = 
\begin{cases}
    L,   & \text{if $M$'s $j$th head moves a block to the \textbf{left} after epoch $i$} \\
    S,  & \text{if $M$'s $j$th head remains in the \textbf{same} block after epoch $i$}\\
    R,   & \text{if $M$'s $j$th head moves a block to the \textbf{right} after epoch $i$.} 
\end{cases}
\end{align*}

A pattern determines the data flow of the computation: for every epoch, it tells us which earlier epoch last worked on the block that the current epoch reads. There are $3^{d(t/b-1)}=2^{O(t/b)}$ patterns, and we do not know a priori which one the computation follows, so we \emph{try all of them}, constructing one circuit-CSP (or 3-CNF) per pattern.\footnote{Strictly speaking, it suffices to only break the runtime up into epochs, without breaking the tapes into blocks. This comes at the cost of increasing the number of head patterns, which worsens our parameters slightly in~\Cref{thm:lossless-general}, but~\Cref{fact:csp} is strong enough that we are still able to recover~\Cref{thm:lossless}. Using~\cite{HPV77} avoids this loss and has the added benefit of making the construction easier to visualize.}

% Highlight colour for the block under each head
\colorlet{headblockcol}{blue!25}
% \colorlet{headtriangle}{gray!50!red}

% Geometry
\def\cw{0.3}     % cell width
\def\ch{0.4}     % cell height
\def\bw{1.2}     % block width (4 cells drawn, standing for b)
\def\rowsep{1.2}  % vertical distance between tapes
% Block positions (left edge of each drawn block):
%   \blkone = B_1,  \blktwo = B_2,  \blklast = B_{t/b}
\def\blkone{0}
\def\blktwo{1.2}
\def\blklast{3.2}
\def\tapelen{4.4} % right end of each tape

% ------------------------------------------------------------------
% \threetapetm{blk1}{cell1}{blk2}{cell2}{blk3}{cell3}{title}
%   To be used inside a tikzpicture (wrap in a scope to shift it).
%   blk   = left edge of the block holding that tape's head
%           (\blkone, \blktwo or \blklast); it is highlighted.
%   cell  = which of the 4 drawn cells (1-4) in that block the head is on.
%   title = label centred above the tapes.
% ------------------------------------------------------------------
\newif\iftapelabels \tapelabelstrue  % \tapelabelsfalse hides the "Tape k" labels
\newcommand{\threetapetm}[7]{%
  % highlighted blocks (drawn first, underneath the grid)
  \fill[headblockcol] (#1, 0)          rectangle ++(\bw, \ch);
  \fill[headblockcol] (#3, -\rowsep)   rectangle ++(\bw, \ch);
  \fill[headblockcol] (#5, -2*\rowsep) rectangle ++(\bw, \ch);
  % tapes
  \foreach \tape in {1,2,3} {
    \pgfmathsetmacro{\y}{-(\tape-1)*\rowsep}
    \foreach \bx in {\blkone, \blktwo, \blklast} {
      \foreach \c in {0,...,3}
        \draw ({\bx + \c*\cw}, \y) rectangle ++(\cw, \ch);
      \draw[very thick] (\bx, \y) rectangle ++(\bw, \ch);
    }
    \draw ({\blktwo+\bw}, \y) -- (\blklast, \y);
    \draw ({\blktwo+\bw}, \y+\ch) -- (\blklast, \y+\ch);
    \node at ({(\blktwo+\bw+\blklast)/2}, \y+\ch/2) {$\cdots$};
    \iftapelabels\node[anchor=east, font=\small] at (-0.08, \y+\ch/2) {Tape \tape};\fi
  }
  % title
  \node[above] at ({\tapelen/2}, \ch+0.05) {#7};
  % heads
  \node[head] at ({#1 + (#2-0.5)*\cw}, -0.16+0.60) {};
  \node[head] at ({#3 + (#4-0.5)*\cw}, -\rowsep-0.16+0.60) {};
  \node[head] at ({#5 + (#6-0.5)*\cw}, -2*\rowsep-0.16+0.60) {};

  %  \node[oldhead] at ({#1 + (#2-0.5)*\cw}, -0.16) {};
  % \node[oldhead] at ({#3 + (#4-0.5)*\cw}, -\rowsep-0.16) {};
  % \node[oldhead] at ({#5 + (#6-0.5)*\cw}, -2*\rowsep-0.16) {};

}

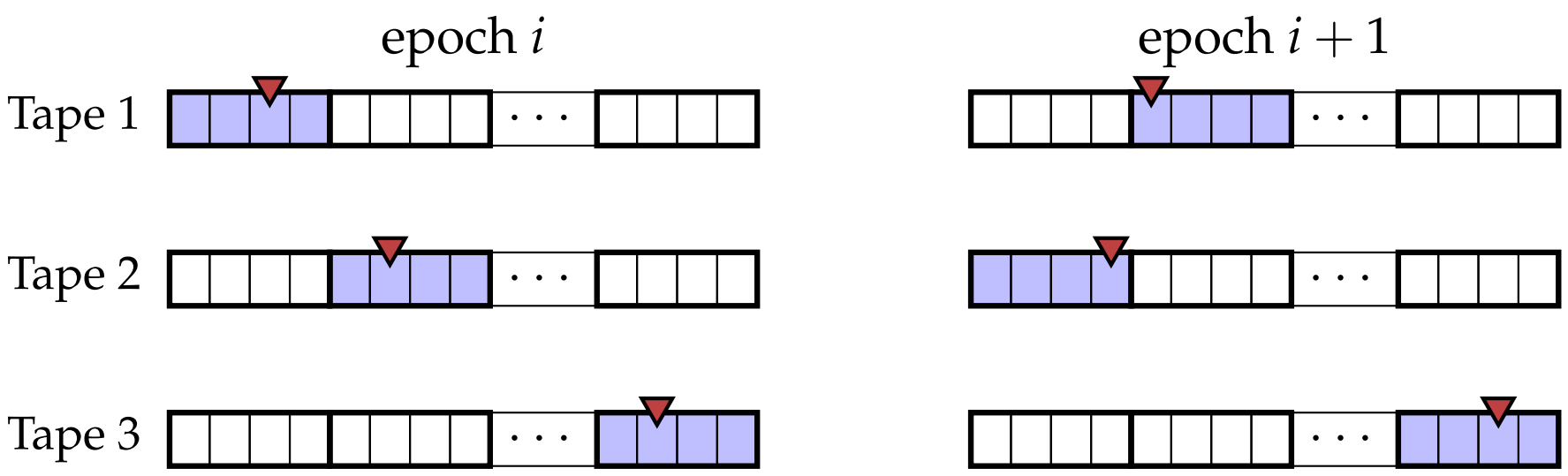
\begin{figure}[ht]
\centering
\begin{tikzpicture}[
    oldhead/.style={ fill=black, isosceles triangle, isosceles triangle apex angle=60,
                   shape border rotate=90, inner sep=0pt, minimum height=0.18cm},
    head/.style = {draw=black, thick, fill=headtriangle, isosceles triangle, isosceles triangle apex angle=60,
                   shape border rotate=270, inner sep=0pt, minimum height=0.20cm},
    move/.style = {-latex, thick},
]
\def\midy{-\rowsep+\ch/2}   % vertical centre of the tapes
\def\rshift{6.0}           % horizontal offset of the right-hand machine

% epoch i: Tape 1 in B_1, Tape 2 in B_2, Tape 3 in B_{t/b}
\threetapetm{\blkone}{3}{\blktwo}{2}{\blklast}{2}{epoch $i$}

% arrow from epoch i to epoch i+1
%\draw[move] (\tapelen+0.2, \midy) -- (\rshift-0.2, \midy);

% epoch i+1: Tape 1 moved right into B_2, Tape 2 moved left into B_1,
%            Tape 3 moved within B_{t/b}
\begin{scope}[xshift=\rshift cm]
  \tapelabelsfalse  % rows line up with the left machine's labels
  \threetapetm{\blktwo}{1}{\blkone}{4}{\blklast}{3}{epoch $i+1$}
\end{scope}
\end{tikzpicture}
\caption{The same 3-tape Turing machine at the start of epochs $i$ and $i+1$; the block that each head stays in during that epoch is shaded. In this example, the pattern for epoch $i$ is $\pi_i = (R, L, S)$: the head on tape~1 moves right one block, the head on tape~2 moves left one block, and the head on
tape~3 remains in the same block.}
\label{fig:pattern}
\end{figure}

\paragraph{How to construct a circuit-CSP from a segmented NTM.}
Let's suppose we have fixed a guess for the pattern $\pi$. Using a result of~\cite{PippengerF79} (restated in~\Cref{lem:pf-nd}), we construct, for each epoch $i$, a constraint circuit $C_i$ of size $O(b\log b)$ that ensures the validity of the computation of $M$ within that epoch. The final circuit-CSP comes from taking the AND of all the $C_i$'s.

The variables of circuit $C_i$ fall into three groups: 
\begin{itemize}
    \item $x^{(i)}$ describes the \emph{block configuration} at the start of epoch $i$ (the block contents, head positions, and state),

    \item $y^{(i)}$ describes the block configuration at the end of epoch $i$, and

    \item  $w^{(i)}$ are the $b$ nondeterministic choices made by $M$ during epoch $i$.
\end{itemize}  
Then, $C_i$ checks whether $y^{(i)}$ is an honest description of the configuration after running $M$ for $b$ steps from configuration $x^{(i)}$ with nondeterministic choices $w^{(i)}$. If it's the last epoch, we of course additionally check whether $M$ accepts. 

The variables $x^{(i)}$ and $y^{(i)}$ are shared among circuits. Each block's content in $x^{(i)}$ is the corresponding part of $y^{(i')}$ for the
last epoch $i'$ that worked on that block, as determined by $\pi$, and the state and head positions come from $y^{(i-1)}$. For example, in~\Cref{fig:hpv-graph}, since epochs $1$ and $4$ work on block $0$, and epochs $2$ and $3$ work on block $1$, the output variables corresponding to block $0$ in $y^{(1)}$ are identified with the inputs for block $0$ in $x^{(4)}$, and similarly, the output variables corresponding to block $1$ in $y^{(2)}$ are identified with the inputs for block $1$ in $x^{(3)}$.

Sharing variables in this way allows us to enforce consistency between epochs. Notice also that each variable is shared among at most 2 circuits. This means the density of our circuit-CSP is at most $2$. See \Cref{fig:hpv-phi} for an illustration. 

% % Figure: HPV computation graph and circuit-CSP for a single tape TM with pi = (R,S,L).
% % Needs only \usepackage{tikz} in the preamble: the TikZ libraries and colours
% % are loaded/defined here, so \input{fig_phi_pi} (or pasting this) is enough.
\usetikzlibrary{arrows.meta,bending,calc}
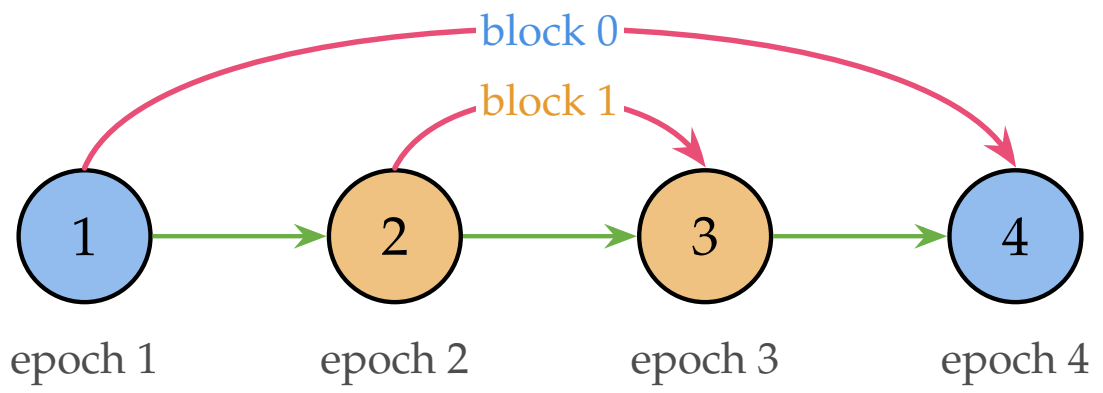
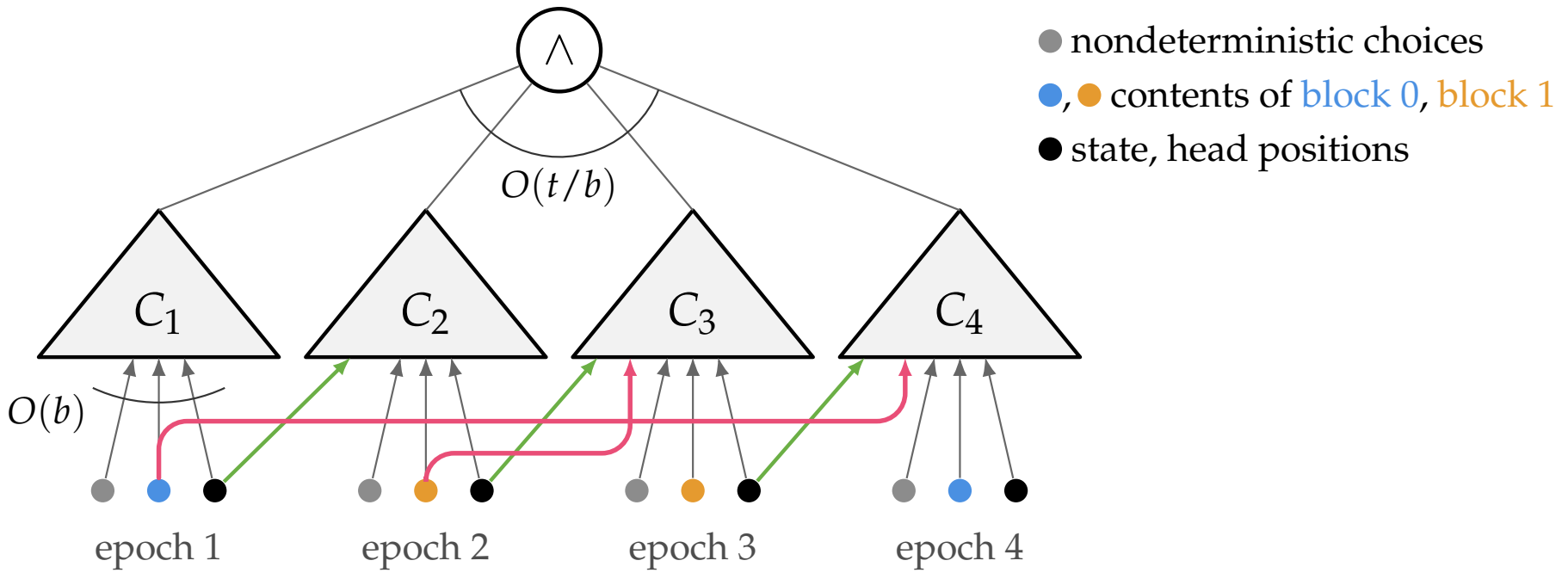
\begin{figure}[!htbp]
\centering
\definecolor{blkA}{HTML}{4A90E2}  % block 0 (blue)
\definecolor{blkB}{HTML}{E59A2F}  % block 1 (orange)
\definecolor{diff}{HTML}{E94E77}  % pink: block contents
\definecolor{purp}{HTML}{6BAF45}  % green: state and head position
% The drawings have the same horizontal extent, so this common width
% gives them the same scale and aligns their epoch columns. The extent is
% symmetric about the epochs (it includes empty space on the left to balance
% the legend on the right), so 0.89\linewidth draws them at the same size as
% the old 0.7\linewidth did.
\newcommand{\hpvDrawingWidth}{0.89\linewidth}

\begin{subfigure}[t]{\linewidth}
\centering
\resizebox{\hpvDrawingWidth}{!}{%
\begin{tikzpicture}[
    x=1cm, y=1cm,
    epoch/.style={circle, draw, thick, minimum size=0.85cm, inner sep=0pt, font=\small, fill=gray!10},
    arr/.style={thick, purp, -{Stealth[length=6pt]}},
    carr/.style={line width=1pt, diff, line cap=round, -{Stealth[length=7pt, bend]}},
    v/.style={circle, fill=black, inner sep=0pt, minimum size=5pt},
  ]
  \def\gap{2.0}   % horizontal distance between epochs (as in fig_phi_pi.tikz)
  \path (-4.6,0) -- ({3*\gap+4.6},0); % common extent, symmetric about the middle epoch column (x = 1.5*\gap) so the drawings sit centred
  % ---- vertices = epochs, shaded by the block the head works in (blue: block 0, orange: block 1)
  \foreach \i/\blk/\bc in {1/0/blkA, 2/1/blkB, 3/1/blkB, 4/0/blkA} {
    \node[epoch, fill=\bc!60] (e\i) at ({\gap*(\i-1)},0) {$\i$};
    \node[font=\scriptsize, black!70, below=3pt] at (e\i.south) {epoch $\i$};
  }
  % ---- state and head positions: i -> i+1
  \foreach \i [evaluate=\i as \j using int(\i+1)] in {1,2,3} \draw[arr] (e\i) -- (e\j);
  % ---- block contents: last epoch on the block -> next epoch on it (depends on pi)
  \draw[carr] (e2.north) .. controls +(0.3,0.6) and +(-0.3,0.6) .. (e3.north);
  \draw[carr] (e1.north) .. controls +(0.5,1.2) and +(-0.5,1.2) .. (e4.north);
  \node[font=\scriptsize, blkB, fill=white, inner sep=1pt] at ($(e2)!0.5!(e3)+(0,0.88)$) {block $1$};
  \node[font=\scriptsize, blkA, fill=white, inner sep=1pt] at ($(e1)!0.5!(e4)+(0,1.33)$) {block $0$};
\end{tikzpicture}%
}
\caption[Computation graph]{Computation graph. Each node represents an epoch, color coded according to the block that contains the tape head during said epoch.
\textcolor{purp}{Green} arrows carry the state and head position to the next epoch.
\textcolor{diff}{Pink} arrows carry a block's contents from its last visit to its next visit.}
\label{fig:hpv-graph}
\end{subfigure}
\par\medskip

\begin{subfigure}[t]{\linewidth}
\centering
\resizebox{\hpvDrawingWidth}{!}{%
\begin{tikzpicture}[
    x=1cm, y=1cm,
    gate/.style={circle, draw, thick, minimum size=0.62cm, inner sep=0pt, font=\small, fill=white},
    v/.style={circle, fill=black, inner sep=0pt, minimum size=5pt},
    w/.style={thin, black!60},
    red/.style={line width=1pt, diff},
  ]
  \def\gap{2.0}   % horizontal distance between epochs
  \def\dx{0.42}   % spacing of the three dots within an epoch
  \path (-4.6,0) -- ({3*\gap+4.6},0); % common extent, symmetric about the middle epoch column (x = 1.5*\gap) so the drawings sit centred
  % ---- variables of epoch i: w^(i), block part of y^(i), state/head part of y^(i)
  \foreach \i/\bc in {1/blkA,2/blkB,3/blkB,4/blkA} {
    \pgfmathsetmacro{\xc}{\gap*(\i-1)}
    \node[v, fill=black!45] (W\i) at (\xc-\dx,0) {};
    \node[v, fill=\bc]      (Y\i) at (\xc,0)     {};
    \node[v]                (S\i) at (\xc+\dx,0) {};
    \node[font=\scriptsize, black!70] at (\xc,-0.45) {epoch $\i$};
  }
  % ---- constraint circuits C_i (triangles, apex up), inputs spread along the base
  \def\hw{0.9}    % half-width of a triangle's base
  \foreach \i in {1,...,4} {
    \pgfmathsetmacro{\xc}{\gap*(\i-1)}
    \coordinate (t\i) at (\xc,1.35);
    \draw[thick, fill=gray!10, line join=round]
      ($(t\i)+(-\hw,-0.35)$) -- ($(t\i)+(\hw,-0.35)$) -- ($(t\i)+(0,0.75)$) -- cycle;
    \node[font=\small] at ($(t\i)+(0,-0.02)$) {$C_{\i}$};
    \coordinate (bl\i) at ($(t\i)+(-\hw,-0.35)$);
    \coordinate (br\i) at ($(t\i)+(\hw,-0.35)$);
    \coordinate (a\i) at ($(t\i)+(0,0.75)$);
  }
  % Wires are straight, with the arrowhead along the line.
  % \own{i}: the circuit's own variables w^(i), y^(i) enter the base nearly above
  %          their dots, leaning slightly towards the centre (see \pull).
  % \ext{i}{m}{style/dot, ...}: the m inputs from earlier epochs enter at evenly spaced
  %          points on the part of the base left of the own wires (list them left to right).
  \def\extL{-0.86}\def\extR{-0.26}   % that part of the base, relative to the circuit's centre
  \def\pull{0.45}   % own wires end at this fraction of their dot's offset from the centre,
                    % so they lean slightly in towards the middle of the circuit
  \def\own#1{%
    \foreach \vv in {W#1,Y#1,S#1}
      \draw[w, -{Latex[length=4pt]}] (\vv) -- ($(t#1 |- bl#1)!\pull!(\vv |- bl#1)$);}
  \def\ext#1#2#3{%
    \foreach \st/\vv [count=\k] in {#3}
      \csname wire\st\endcsname{\vv}{$(t#1)+({\extL+(\extR-\extL)*(\k-0.5)/#2},-0.35)$};}
  % grey wires are straight
  \def\wirew#1#2{\draw[w, -{Latex[length=4pt]}] (#1) -- (#2);}
  % purple wires: state and head positions from y^(i-1), the purple edges of (a)
  \def\wirepur#1#2{\draw[purp, line width=0.8pt, -{Latex[length=4pt]}] (#1) -- (#2);}
  % red wires are smooth S-curves that only ever go up and to the right: with
  % horizontal distance dx and vertical distance dy, the control points are at
  % start + (\sx dx, \sy dy) and end - (\sx dx, \sy dy). Since \sy < 1/2 the control
  % heights are in order, so the curve never turns back down; offsets that scale
  % with dx spread the bend along the whole wire instead of bunching it at the ends.
  % The horizontal offset is capped at \smax, so long wires still leave their dot and
  % enter their circuit at a reasonable slope.
  \def\sx{0.22}\def\sy{0.46}\def\smax{0.7cm}
  \def\wirered#1#2{\draw[red, line cap=round, -{Latex[length=4pt, bend]}]
    let \p1=(#1), \p2=(#2), \n1={min(\sx*(\x2-\x1),\smax)}, \n2={\sy*(\y2-\y1)} in
    (#1) .. controls ($(\p1)+(\n1,\n2)$) and ($(\p2)-(\n1,\n2)$) .. (\p2);}
  % pink wires drawn as vertical -- horizontal -- vertical, with rounded corners:
  % up from the dot to a fixed height, across, then straight up into the circuit.
  % The long wire (epoch 1 -> C_4) runs at height \hyL, the short one nested inside
  % it (epoch 2 -> C_3) lower, at \hyS, as in panel (a).
  \def\hyL{0.52}\def\hyS{0.28}\def\rc{6pt}
  \def\elbow#1#2#3{\draw[red, line cap=round, rounded corners=\rc, -{Latex[length=4pt]}]
    (#2) -- ($(#2)+(0,#1)$) -| (#3);}
  \def\wireredL#1#2{\elbow{\hyL}{#1}{#2}}
  \def\wireredS#1#2{\elbow{\hyS}{#1}{#2}}
  \foreach \i in {1,...,4} {\own{\i}}
  % state and heads from y^(i-1); red: block contents from y^(i'), i' the last epoch on that block
  % \extat{i}{k}{m}{style}{dot}: one incoming wire into slot k (of m, left to right)
  % of circuit C_i, so the slots can be chosen independently of drawing order
  \def\extat#1#2#3#4#5{\csname wire#4\endcsname{#5}{$(t#1)+({\extL+(\extR-\extL)*(#2-0.5)/#3},-0.35)$}}
  % green wires first (left-hand slot), so they sit behind the pink ones (right-hand slot)
  \extat{2}{1}{1}{pur}{S1}
  \extat{3}{1}{2}{pur}{S2}
  \extat{4}{1}{2}{pur}{S3}
  \extat{3}{1.8}{2}{redS}{Y2}   % slightly left of the usual right-hand slot
  \extat{4}{2}{2}{redL}{Y1}
  % ---- fan-in of a circuit (shown on C_1; the same O(b) for every C_i)
  \coordinate (fc) at ($(t1)+(0,0.48)$);   % centre of a flat arc just below C_1's base
  \draw[thin, black!80] ($(fc)+(245:1.17)$) arc[start angle=245, end angle=295, radius=1.17];
  \node[font=\scriptsize, anchor=north east, inner sep=1pt] at ($(fc)+(245:1.17)$) {$O(b)$};
  % ---- AND over the epochs
  \node[gate] (AND) at ({1.5*\gap},3.3) {$\wedge$};
  \foreach \i in {1,...,4} \draw[w] (AND) -- (a\i);
  % fan-out arc
  \pgfmathsetmacro{\aL}{atan2(1.2,-1.5*\gap)}   % angle towards a1
  \pgfmathsetmacro{\aR}{atan2(1.2, 1.5*\gap)}   % angle towards a4
  \draw[thin, black!80] ($(AND.center)+(180+\aR:0.8)$) arc[start angle=180+\aR, end angle=360-\aR, radius=0.8];
  \node[font=\scriptsize, fill=white, inner sep=1pt, below=1pt] at ($(AND.center)+(270:0.8)$) {$O(t/b)$};
  % ---- legend (short), in the top-right corner of the panel
  \node[font=\scriptsize, anchor=north east, align=left] at ({3*\gap+4.6},3.62)
    {\tikz\node[v, fill=black!45]{}; nondeterministic choices\\[2pt]
     \tikz\node[v, fill=blkA]{};,\,\tikz\node[v, fill=blkB]{}; contents of {\color{blkA}block $0$}, {\color{blkB}block $1$}\\[2pt]
     \tikz\node[v]{}; state, head positions};
\end{tikzpicture}%
}
\caption[Circuit-CSP]{Circuit-CSP $\Phi_\pi$. The variables below $C_i$ encode epoch $i$'s
nondeterministic choices (\textcolor{black!45}{gray}), final block contents
(\textcolor{blkA}{blue} or \textcolor{blkB}{orange}), and final state and head position
(black). The \textcolor{purp}{green} and \textcolor{diff}{pink} wires supply inputs
from earlier epochs as specified by the computation graph.}
\label{fig:hpv-csp}
\end{subfigure}

\caption{A single-tape Turing machine with head pattern $\pi=(R,S,L)$ and four epochs.}
\label{fig:hpv-phi}
\end{figure}

\paragraph{How to construct a 3-CNF formula from a circuit-CSP.} We can convert each constraint circuit $C_i$ in the previous step into a 3-CNF formula over $O(b\log b)$ variables using a result of~\cite{Tseitin68} (\Cref{lem:tse}). There are $O(t/b)$ different constraint circuits, so taking the AND of all of them results in a 3-CNF over $O(t\log b)$ variables. 

\paragraph{Summary.} You can view our argument as consisting of the same two steps as Cook--Levin just applied to segments of the Turing machine execution instead of the whole thing. We summarize the argument here:
\begin{enumerate}
    \item Divide the Turing machine's runtime into epochs, applying~\cite{HPV77} to show it can be made block-respecting. 
    Each pattern $\pi$ for how the heads move between blocks at the boundaries of each epoch will give rise to a different circuit-CSP. 
    \item For a fixed pattern $\pi$, use~\cite{PippengerF79} to convert each epoch into a constraint circuit. Taking the AND of all these circuits results in a circuit-CSP. 
    \item Use~\cite{Tseitin68} to convert each circuit-CSP into a 3-CNF. 
\end{enumerate}

\noindent\Cref{lem:cl-for-CSP} follows from just steps $1$ and $2$. Step $3$ is only needed for obtaining~\Cref{cor:cl}.

\section{Preliminaries}

\paragraph{Notation.} We write $[n]=\{1,\ldots, n\}$ and $[0,n]=\{0,\ldots,n\}$. All functions $f: \N \rightarrow \N$ are assumed to be growth functions (i.e.~time-constructible and non-decreasing) unless stated otherwise. All $\log$'s are base 2.

\paragraph{Block-respecting Turing machines.} The proof of~\Cref{thm:lossless-general} uses the notion of \emph{block-respecting Turing machines} from Hopcroft, Paul, and Valiant~\cite{HPV77}. Roughly speaking, a $b$-block-respecting Turing machine has each tape divided into blocks of size $b$, and tape heads are only allowed to cross from one block to another at time steps that are integer multiples of $b$ (so at most one block crossing every $b$ steps). Intuitively, assuming a block-respecting Turing machine helps our proof since we can then think of its computation on a given block as a constraint of arity $O(b)$ over variables representing the content of the $b$ tape-cells of the block. 
Concretely, we use the following slight modification of~\cite{HPV77}:

\begin{lemma}[NTMs can be made block-respecting~\cite{HPV77}]\label{lem:hpv}
    Let $M$ be a $d$-tape nondeterministic Turing machine running in time $\timefn(n)$.  Then, for every function $b \colon \mathbb{N} \to \mathbb{N}$ with $b(n) \leq \timefn(n)$, there exists a nondeterministic Turing machine $M'$ that accepts the same language as $M$. Furthermore, $M'$ uses $O(d)$ tapes, runs in time $O(\timefn(n))$, and is $b(n)$-block-respecting when receiving an input of length $n$.
\end{lemma}

\paragraph{Converting Turing machines into circuits.}
We will also use a form of Pippenger--Fischer's result on compiling deterministic Turing machines into circuits~\cite{PippengerF79}. To state it we will need some notation. Given a $d$-tape deterministic Turing machine $M$ over an alphabet $\Gamma$, and set of states $Q$, we define its set of \emph{block configurations}  as $\mathcal{C}_{M,b} :=(\Gamma^b)^d \times Q \times [0,b+1]^d$ for $b \geq 1$, representing the content of a size-$b$ block on each of the $d$ tapes, together with a state, and an offset describing the starting position of each head within the block (the positions $0$ and $b+1$ indicate that the head is just to the left or just to the right of the block, respectively). Then, a \emph{block computation} is a function
\[
\mathcal{B}_{M,b}\colon \mathcal{C}_{M,b} \to \mathcal{C}_{M,b}\cup  \{ \bot \}
\]
which maps a block configuration to the one resulting after $b$ steps of computation by $M$ if all heads stayed within their block in the first $b-1$ steps of computation, and to $\bot$ if some head exited its block in the first $b-1$ steps or began in position $0$ or $b+1$.
Informally, the result of Pippenger--Fischer says $\mathcal{B}_{M,b}$, interpreted as a boolean function, can be computed by a circuit with $O(b \log b)$ gates. More formally,
let $ L_{M,b} := db \lceil\log|\Gamma|\rceil  + \lceil\log |Q|\rceil + d \lceil\log (b+2)\rceil$ be the number of bits required to describe a block configuration, and let
$\mathsf{enc}_{M,b} \colon \mathcal{C}_{M,b} \to \{0,1\}^{L_{M,b}}$ be a function that encodes block configurations in binary in the natural way. Then, we can state Pippenger--Fischer's result as follows.

\begin{lemma}[Converting TMs into Circuits \cite{PippengerF79}]
    Let $M$ be a $d$-tape deterministic Turing machine, and let $b \geq 1$ be an integer. Let $f_{M,b} \colon \{0,1\}^{2L_{M,b}}\to \{0,1\}$ be such that $f_{M,b}({x}, {y}) = 1$ for ${x}, {y}\in \zo^{L_{M,b}}$ if and only if ${x}$ and ${y}$ are encodings of block configurations $\beta_1$ and $\beta_2$, respectively, and $\mathcal{B}_{M,b}(\beta_1) = \beta_2$. Then a circuit for $f_{M,b}$ with $O(b \log b)$ gates can be computed deterministically in time $\poly(b)$.
\end{lemma}

There is a natural analogue of the above for nondeterministic Turing machines. We assume that a nondeterministic machine $M$ makes one binary choice per step, or equivalently, it can read the $i$th bit of a witness $w$ during the $i$th step. For a block configuration $\beta \in\mathcal{C}_{M,b}$ and a witness
$w\in\{0,1\}^b$, let $\mathcal{B}_{M,b}(\beta,w)\in\mathcal{C}_{M,b}\cup\{\bot\}$ be
defined exactly as $\mathcal{B}_{M,b}(\beta)$ in the deterministic case except now it additionally takes in $w$ as input. This gives a \emph{nondeterministic block computation}
\[
  \mathcal{B}_{M,b}\colon \mathcal{C}_{M,b}\times\{0,1\}^b \to \mathcal{C}_{M,b}\cup\{\bot\}.
\]

\begin{lemma}[Converting NTMs into circuits~\cite{PippengerF79}]\label{lem:pf-nd}
  Let $M$ be a $d$-tape nondeterministic Turing machine, and let $b\ge1$ be an
  integer. Let $g_{M,b}\colon\{0,1\}^{2L_{M,b}+b}\to\{0,1\}$ be such that
  $g_{M,b}( x, w, y)=1$ for $ x, y\in\{0,1\}^{L_{M,b}}$ and
  $ w\in\{0,1\}^b$ if and only if ${x}$ and ${y}$ are encodings of block configurations $\beta_1$ and $\beta_2$, respectively, and  $\mathcal{B}_{M,b}(\beta_1, w)=\beta_2$.
  Then a circuit for $g_{M,b}$ with $O(b\log b)$ gates can be computed
  deterministically in time $\poly(b)$.
\end{lemma}

\paragraph{Converting circuits into 3-CNFs.} We will also need a way of converting circuits into 3-CNF formulas. This result was first discovered by~\cite{Tseitin68}, although it's been rediscovered in different forms many times since. 

\begin{lemma}[Converting circuits into CNFs~\cite{Tseitin68}] \label{lem:tse} Let $C({x})$ be a circuit over $n$ variables with $g$ gates of fan-in at most $2$. There is a 3-CNF formula $\varphi({x}, {y})$ over $n+g$ variables and $O(g)$ clauses such that for every assignment to the variables ${x}$ we have $C({x}) = 1$ if and only if $\exists {y}\, \varphi({x}, {y}) = 1$. Furthermore $\varphi$ can be computed in time $\poly(n+g)$. 
    
\end{lemma}

\section{Proofs of~\Cref{thm:lossless-general} and~\Cref{thm:lossy-general}} \label{sec:lossless-thm-proofs}

\begin{proof}[Proof of~\Cref{thm:lossless-general} assuming~\Cref{lem:cl-for-CSP}]
Let $M$ be a nondeterministic Turing machine with running time $t(n)$, and $k(n)$ be the arity function in the hypothesis.
Given an input $x$ of length $n$,
we will deterministically decide whether $M$ accepts $x$ as follows.
First, we use the algorithm from~\Cref{lem:cl-for-CSP} on $M$, $x$, and $b(n) := \lfloor c \cdot k(t(n))\rfloor$ for some sufficiently small $0 < c < 1$,\footnote{Since we always have $k(n) \leq n$, it follows that $b(n) \leq t(n)$. } thus obtaining a collection of circuit-CSPs $\Phi_1\ldots\Phi_m$, each with $N$ variables, where  $t(n) \leq N \leq Ct(n)$ for some fixed $C > 1$. This step takes time $m\cdot\poly(t(n))$. 

The assumption of~\Cref{thm:lossless-general} tells us that each of these circuit-CSPs $\Phi_\ell$ with $O(t(n))$ variables, arity $O(b(n)) \leq k(t(n)) \leq k(N)$, and density at most $2$ can be solved in time $2^{O(t(n)/s(t(n)))}\cdot\poly(|\Phi_\ell|)=2^{O(t(n)/s(t(n)))}$. There are $m$ such Circuit-CSPs, so solving all of them takes time $m\cdot 2^{O(t(n)/s(t(n)))}$. 
We output that $M$ accepts $x$ iff any $\Phi_{\ell}$ is satisfiable. Condition~\ref{item:csp-correct} of~\Cref{lem:cl-for-CSP} ensures the correctness of this algorithm. Plugging in $m=2^{O(t(n)/b(n))}$, the total runtime is 
\begin{align*}
    2^{O(t(n)/b(n))}\big(\poly(t(n)) +  2^{O(t(n)/s(t(n)))}\big) =  2^{O(t(n)/b(n))}\cdot 2^{O(t(n)/s(t(n)))}
\end{align*}
since $s(t(n))\leq t(n)/\log t(n)$ by hypothesis. 
\end{proof}

\begin{proof}[Proof of~\Cref{thm:lossy-general} assuming~\Cref{cor:cl}] The proof is nearly identical to that of~\Cref{thm:lossless-general}. Given a nondeterministic TM $M$ running in time $t := \timefn(n)$ and an input $x$ of size $n$, we compute $M(x)$ as follows. First, we run the reduction in~\Cref{cor:cl} to obtain a collection of 3-CNF formulas $\varphi_1\ldots\varphi_m$. This step takes time $m \cdot \poly(t(n))$. Then, since each $\varphi_\ell$ has $O(t \log b)$ variables and clauses, the assumption of~\Cref{thm:lossy-general} tells us that each 3-CNF can be solved in time $2^{O(t\log b/s(t \log b))}$, so solving \emph{all} of them takes time $m\cdot 2^{O(t\log b/s(t\log b))}$. We accept iff any $\varphi_{\ell}$ is satisfiable. Plugging in $m=2^{O(t/b)}$ makes the total runtime $2^{O(t/b)}\cdot 2^{O(t\log b/s(t\log b))}.$ Taking $b:=s(t(n))$ makes this 
\begin{align*}
    2^{O(t/s(t))}2^{O(t\log s(t)/s(t\log s(t)))} =  2^{O(t\log s(t) / s(t))}
\end{align*}
since $s$ is non-decreasing. 
\end{proof}

\section{Proof of~\Cref{lem:cl-for-CSP} and~\Cref{cor:cl}}

\begin{proof}[Proof of~\Cref{lem:cl-for-CSP}]
    Fix $M, t,$ and $b$. First, by~\Cref{lem:hpv}, there is a $b$-block-respecting Turing machine $M'$ accepting the same language as $M$, with $D = O(1)$ tapes. 
    Fix an input $x$ of length $n$, and let $T := c\cdot t(n)$ for some $c > 0$ be an upper bound on the runtime of $M'$ on $x$, which we can easily compute since $t$ is time-constructible. Then, let $\hat{M}$ be a copy of $M'$ in which accepting and rejecting states are made ``absorbing'' (i.e., once reached, the machine remains in the same state, preserving the content of the tapes). This way, $\hat{M}$ runs forever on $x$, but its state is accepting at time $b\lceil T / b \rceil$ if and only if $M$ accepts on $x$.\footnote{This is just for convenience, since this way we do not have to worry about the computation finishing in the middle of an epoch.} For ease of notation, let $r := \lceil T / b\rceil$, so we wish to represent the computation of $\hat{M}$ on $x$ for $r$ epochs of $b$ steps each.
    We enumerate all \emph{patterns} (see~\Cref{subsec:overview}) $\pi \in \{L, S, R\}^{(r-1)\times D}$, and since there are $3^{(r-1)D} = 2^{O(t/b)}$ of them, it suffices to show how to construct a given circuit-CSP $\Phi_\pi$ corresponding to whether $\hat{M}$ is in an accepting state at time $rb$ having followed pattern $\pi$.

    Let $\pi_{i, j} \in \{L, S, R\}$ denote the block movement at the end of epoch $i$ for tape $j$, and let $a_{i, j}$ be the index of the block of the head of tape $j$ during epoch $i$. We thus have 
    $a_{1, j} = 0$ for $j \in [D]$, and  $a_{i+1, j} = a_{i, j} + \delta(\pi_{i, j})$, where $\delta$ maps $L \to -1, S \to 0, R \to 1$.

    For each epoch $i$, let $\Xi_i$ be the circuit obtained from~\Cref{lem:pf-nd} computing $g_{\hat{M}, b}({x}^{(i)}, {w}^{(i)}, {y}^{(i)})$. Our goal is to combine the different $\Xi_i$ in order to obtain $\Phi_\pi$, and the main challenge is the interaction between their variables (e.g., the final content of tape $j$ on a given block after a given epoch needs to match the initial content of a future epoch working on the same block).
   In order to do this formally, we will use a notion of \emph{formal substitution}. First, note that we are free to ``hardwire" constant values into circuits by using \emph{constant gates} labeled with $0$ or $1$; these can be obtained trivially as $(x \land \neg x)$ and $(x \lor \neg x)$ for any variable $x$.
    Now, we think of circuits as directed graphs whose vertices are labeled either as variables or gates. Under this perspective, we start with a disconnected graph where each component is one of the $\Xi_i$, and we will then merge them into a single circuit.
    Concretely, let us define a \emph{formal substitution} as follows: let $C$ be a boolean circuit, whose set of input variables is $V$; then, a formal substitution is simply a function $\sigma \colon V \to W \cup \{0, 1\}$  where $W$ is some set of boolean variables.  In turn, $C[\sigma]$ is the circuit obtained by simultaneously replacing each input variable $v \in V$ by $\sigma(v)$.  In terms of the graph, \Cref{fig:formal-substitution} illustrates the effect of formal substitution.

    \begin{figure}[tbp]
  \centering
  \begingroup
  \definecolor{subBlue}{HTML}{4A90E2}
  \definecolor{subPink}{HTML}{E94E77}
  \definecolor{subOrange}{HTML}{E59A2F}
  \tikzset{
    subpicture/.style={x=1cm,y=1cm,font=\small},
    subinput/.style={circle,draw=none,fill=black!55,
      minimum size=4pt,inner sep=0pt,outer sep=.6pt},
    subshared/.style={subinput,fill=subBlue},
    subgate/.style={circle,draw=black,fill=white,line width=.45pt,
      minimum size=5.8mm,inner sep=0pt},
    subconstant/.style={subgate,draw=subOrange,text=subOrange,
      minimum size=4.8mm,line width=.6pt},
    subwire/.style={-{Stealth[length=1.4mm,width=1mm]},
      draw=black!90,line width=.4pt},
    subhighlight/.style={subwire,draw=subPink,line width=.7pt},
    subfixedwire/.style={subwire,draw=subOrange,line width=.7pt}
  }

  \begin{subfigure}[t]{4.4cm}
    \centering
    \vspace{0pt}
    \begin{tikzpicture}[subpicture]
      % Equal heights keep the two drawings and their captions aligned.
      \path[use as bounding box] (-.35,-.55) rectangle (3.9,2.7);

      \node[subinput,fill=subOrange,
        label={[text=subOrange]below:{$a$}}] (a1) at (0,0) {};
      \node[subshared,label={[text=subBlue]below:{$b$}}]
        (b1) at (1.1,0) {};
      \node[subgate] (and1) at (.55,1.65) {$\land$};
      \draw[subfixedwire] (a1) -- (and1);
      \draw[subhighlight] (b1) -- (and1);
      % \draw[subwire] (and1) -- (.55,2.15) node[above] {$C_1$};

      \node[subshared,label={[text=subBlue]below:{$x$}}]
        (x1) at (2.45,0) {};
      \node[subinput,label=below:{$c$}] (c1) at (3.55,0) {};
      \node[subgate] (not1) at (2.45,.8) {$\neg$};
      \node[subgate] (or1) at (3,1.65) {$\lor$};
      \draw[subhighlight] (x1) -- (not1);
      \draw[subwire] (not1) -- (or1);
      \draw[subwire] (c1) -- (or1);
      % \draw[subwire] (or1) -- (3,2.15) node[above] {$C_2$};
    \end{tikzpicture}
    \caption{Before substitution.}
    \label{fig:substitution-before}
  \end{subfigure}%
  \hspace{.3cm}%
  \begin{minipage}[t]{2cm}
    \centering
    \vspace{0pt}
    \begin{tikzpicture}[subpicture]
      \path[use as bounding box] (-.1,-.55) rectangle (1.9,2.7);
      \draw[-{Stealth[length=2mm,width=1.3mm]},line width=.6pt]
        (.15,1.1) -- (1.65,1.1)
        node[midway,above=3mm,align=center]
          {$\sigma(x)=b$\\$\sigma(a)=1$}
        ;
    \end{tikzpicture}
  \end{minipage}%
  \hspace{.3cm}%
  \begin{subfigure}[t]{4.4cm}
    \centering
    \vspace{0pt}
    \begin{tikzpicture}[subpicture]
      \path[use as bounding box] (-.35,-.55) rectangle (3.9,2.7);
      \node[subconstant] (one2) at (0,0) {$1$};
      \node[subshared,label={[text=subBlue]below:{$b$}}]
        (b2) at (1.25,0) {};
      \node[subinput,label=below:{$c$}] (c2) at (3.3,0) {};
      \node[subgate] (and2) at (.55,1.65) {$\land$};
      \node[subgate] (not2) at (2.2,.8) {$\neg$};
      \node[subgate] (or2) at (2.75,1.65) {$\lor$};
      \draw[subfixedwire] (one2) -- (and2);
      \draw[subhighlight] (b2) -- (and2);
      \draw[subhighlight] (b2) -- (not2);
      \draw[subwire] (not2) -- (or2);
      \draw[subwire] (c2) -- (or2);
    \end{tikzpicture}
    \caption{After substitution.}
    \label{fig:substitution-after}
  \end{subfigure}
  \par
  \endgroup

  \caption{Formal substitution on circuit graphs.
    The substitution $\sigma(x)=b$ and $\sigma(a)=1$, fixing $b,c$.}
  \label{fig:formal-substitution}
\end{figure}
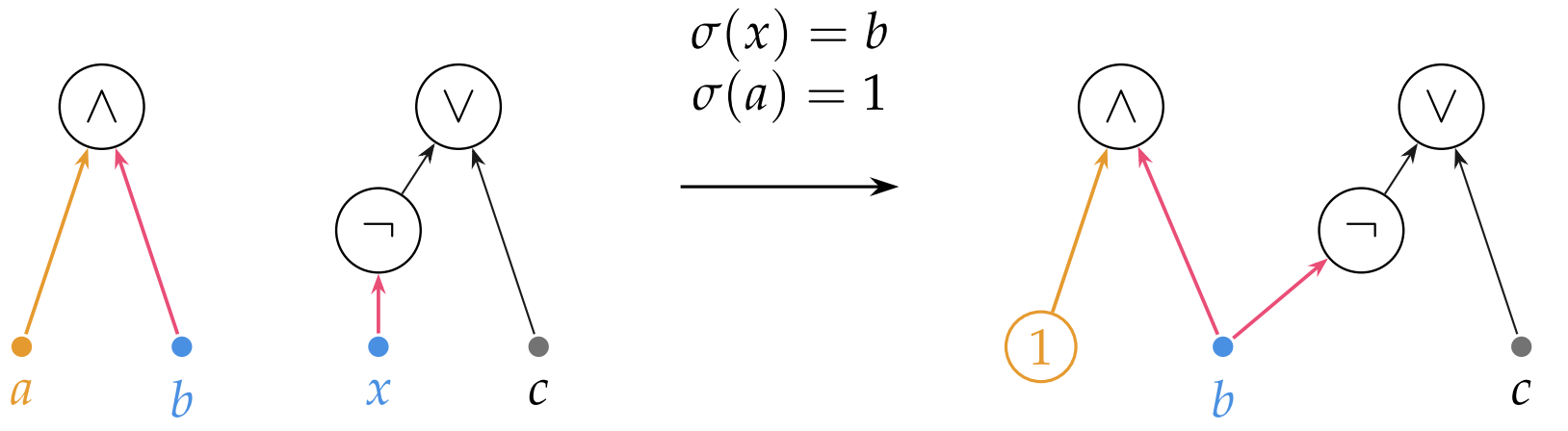

    Now, let us look more closely at the variables involved. Let $\Gamma$ and $Q$ be the alphabet and states of $\hat{M}$, respectively. We can write the input and output configurations ${x}^{(i)}$ and ${y}^{(i)}$ as 
    \[
    {x}^{(i)} = ({u}_{i, 1}, \ldots, {u}_{i, D}, {q}_i, {h}_{i,1}, \ldots, {h}_{i, D}) \quad \text{ and } \quad
    {y}^{(i)} = ({u}'_{i, 1}, \ldots, {u}'_{i, D}, {q}'_i, {h}'_{i,1}, \ldots, {h}'_{i, D}),
    \]
where:
\begin{itemize}
    \item ${u}_{i, j}$ consists of $b \lceil \log |\Gamma|\rceil$ bits representing the content of block $a_{i, j}$ before epoch $i$, and ${u}'_{i, j}$ is analogous but for after epoch $i$.
    \item ${q}_i$ consists of $\lceil\log |Q|\rceil$ bits representing the state before epoch $i$, and ${q}'_i$ the state after epoch $i$; and
    \item ${h}_{i, j}$ consists of $\lceil\log(b+2)\rceil$ bits representing the initial head offset relative to block $a_{i, j}$, and ${h}'_{i, j}$ the final head offset.
\end{itemize}
Next, we will define a formal substitution $\sigma^\star$ that intuitively represents giving the correct input information to $\Xi_i$ based on the results of previous epochs. It will be convenient to define two pieces of notation.
First, for every epoch $i$ and tape $j$, let $\prev(i, j)$ be the last epoch before $i$ that visited block $a_{i, j}$ (or $0$ in case that block had not been previously visited). More formally,
    \[
    \prev(i, j) := \max(\{0\} \cup \{h \in [i-1] \mid a_{h, j} = a_{i, j}\}). 
    \]
Second, for each block index $a_{i, j}$, let $I_j(a_{i, j}) \in \{0,1\}^{\lceil \log |\Gamma| \rceil b}$ be the constant binary encoding of the initial content of that block on tape $j$ (e.g., on every tape except the input tape, it will simply consist of the constant encoding of the blank symbol).
With this notation, we set
\[
\sigma^\star({u}_{i, j}) = \begin{cases}
    I_j(a_{i, j}) & \text{if } \prev(i, j) = 0\\
    {u}'_{\prev(i,j), j} & \text{otherwise}.
\end{cases}
\]
That is, the initial content of a block comes either from the original state of the tapes, or from the output of the last time the machine touched that block.
Similarly, we set
\[
\sigma^\star({q}_i) := \begin{cases}
    \lfloor {q}_{\text{init}} \rfloor & \text{if } i = 1\\
    {q}'_{i-1} & \text{otherwise},
\end{cases}
\]
where $\lfloor {q}_{\text{init}} \rfloor$ is the constant binary encoding for the initial state of $\hat{M}$. So this substitution represents that the state at the beginning of an epoch is either the initial state or the one resulting from the previous epoch.

For the offsets of the heads, the substitution is a bit more cumbersome.   For an integer $v \in [0, b+1]$, let $\lfloor v \rfloor$ denote its binary encoding on $\lceil \log(b+2)\rceil$ bits.
Then, for the first epoch, we have
$\sigma^\star({h}_{1, j}) = \lfloor1\rfloor$, since each head starts at the beginning of its corresponding block. For the epochs after the first, we have three cases:
\begin{enumerate}
    \item If $\pi_{i, j} = S$, then $\sigma^\star({h}_{i+1, j}) = {h}'_{i, j}$,
    \item if $\pi_{i, j} = L$, then
        $\sigma^\star({h}_{i+1, j}) = \lfloor b \rfloor$, and
    \item if $\pi_{i, j} = R$, then
        $\sigma^\star({h}_{i+1, j}) = \lfloor 1 \rfloor$.
\end{enumerate}
Furthermore, we need some consistency checks for the movement of the heads, defined as follows. If $\pi_{i, j} = S$, then the check $E_{i, j}$ is a circuit ensuring that ${h}'_{i, j}$ is indeed a number between $1$ and $b$. If $\pi_{i, j} = L$, then $E_{i, j}$ ensures that ${h}'_{i,j} = \lfloor 0 \rfloor$, and if $\pi_{i, j} = R$, then $E_{i, j}$ ensures that ${h}'_{i, j} = \lfloor b+1 \rfloor$.
Note that each $E_{i, j}$ can be implemented with $O(\log b)$ gates.

Our constraint for every epoch $i < r$ is 
\[
C_i := \Xi_i[\sigma^\star] \land \bigwedge_{j=1}^D E_{i, j}.
\]
For the final epoch $i=r$, let $\Omega_i$ be a circuit whose input variables are ${q}'_i$ and that returns $1$ if ${q}'_i$ equals the binary encoding of any of the accepting states from the set $F \subseteq Q$. Then,
\[
C_i := \Xi_i[\sigma^\star] \land \Omega_i,
\]
and note that $\Omega_i$ uses only a constant number of gates, since $|Q| = O(1)$.
Finally, we simply output $\Phi_\pi := \bigwedge_{i=1}^r C_i$.

The fact that $M$ accepts input $x$ if and only if some $\Phi_\ell$ is satisfiable holds essentially by construction. Each $C_i$ has $O(b)$ variables, and since there are $r = O(t/b)$ of them, the total number of variables is $O(t).$  The fact that each $C_i$ has $O(b \log b)$ gates follows from the fact that each $\Xi_i$ has $O(b \log b)$ gates per~\Cref{lem:pf-nd}, and the $E_{i, j}$ have $O(\log b)$ gates each.
Moreover, each variable appears in at most two constraints: state variables ${q}_i/{q}_i'$ and head offset variables are shared between consecutive epochs, and the variables representing tape contents at the end of an epoch can only be shared with the next epoch touching the corresponding block. Thus, the density is at most $2$, and we conclude the proof.
\end{proof}

\begin{proof}[Proof of~\Cref{cor:cl}]
    By~\Cref{lem:cl-for-CSP}, we obtain circuit-CSPs $\Phi_1, \ldots, \Phi_m$ such that $M$ accepts $x$ if and only if one of them is satisfiable.  
    For each such $\Phi_\pi$,  we 
    iterate over its constraints: 
    $(S_1, C_1), \ldots, (S_r, C_r)$.  Each $C_i$ has $O(b)$ variables and  $O(b \log b)$ gates, so by~\Cref{lem:tse}, we can convert it into a 3-CNF formula $\psi_i(S_i, {z}^{(i)})$, where the additional variables ${z}^{(i)}$ are disjoint across different $i$.
    Then, let $\varphi_\pi := \bigwedge_{i=1}^r \psi_i(S_i, {z}^{(i)})$, and note it has $O(t/ b) \cdot O(b \log b) = O(t \log b)$ variables and clauses.
    For correctness, 
    assume a variable assignment ${a}$ satisfies $\Phi_\pi$, then $C_i({a}|_{S_i}) = 1$ for every $i$, so each
$\psi_i$ has a satisfying extension ${z}^{(i)}$; since the ${z}^{(i)}$ are disjoint,
these combine into a satisfying assignment of $\varphi_\pi$. Conversely, if
$({a}, {z}^{(1)}, {z}^{(2)}, \dots)$ satisfies $\varphi_\pi$, then each $\psi_i$ is
satisfied, so $C_i({a}|_{S_i}) = 1$ for all $i$ and ${a}$ satisfies $\Phi_\pi$.
\end{proof}

\section{Proof of~\Cref{fact:csp}}

We prove~\Cref{fact:csp} assuming~\cite{openai26}'s Theorem D.1. To state~\cite{openai26}'s result, we need the following definition:

\begin{definition}[Boolean tensor network]\label{def:tensor}
Let $R$ be a commutative ring. A \emph{Boolean tensor network} over $R$ consists of a finite bipartite graph $G=(V,E)$, in which parallel edges are allowed, together with a function
\[
  T_v\colon\{0,1\}^{\delta(v)}\to R
\]
for each vertex $v$, where $\delta(v)$ is the set of edges incident to $v$. Think of each edge $e$ as a Boolean variable $x_e$, and of $T_v$ as a weight that depends only on the variables of the edges at $v$. The \emph{contraction} of the network is the sum, over all ways of assigning a bit to every edge, of the product of the weights of all vertices:
\[
  Z=\sum_{x\in\{0,1\}^{E}}\ \prod_{v\in V}T_v\bigl(x|_{\delta(v)}\bigr),
\]
where $x|_{\delta(v)}$ is the restriction of $x$ to the edges at $v$. The \emph{degree} of the network is the maximum degree of $G$. 
\end{definition}
 
\noindent We can now state a slightly simplified version of~\cite{openai26}'s Theorem D.1: 

\begin{theorem}[{\cite[Theorem~D.1]{openai26}}]\label{thm:d1}
Let $\sigma\ge 2$, and let $L$ be the description length of a Boolean tensor network with $O(\sigma)$ edges, at most $\poly(\sigma)$ vertices and maximum degree at most $\lceil\sigma^{9/10}\rceil$. Let every $T_v$ map to the ring $R=\mathbb Z/2^{r}\mathbb Z$ for some $1\leq r\leq O(\sigma)$. Suppose each value of $T_v$ can be computed from the description in time $\poly(\sigma+L)\cdot 2^{O(\sigma^{9/10})}$ and space $\poly(\sigma+L)$. Then the contraction can be computed deterministically in time $\poly(\sigma+L) \cdot 2^{O(\sigma/\sqrt{\log\sigma})}$.
\end{theorem}

\noindent We are now ready to prove that~\Cref{fact:csp} follows from~\Cref{thm:d1}:

\begin{proof}[Proof of~\Cref{fact:csp}]
Let $\Phi$ be a circuit-CSP with $n$ variables, arity at most $n^{9/10}$ and constant density. We build a tensor network whose contraction is $\#\Phi$ and then apply Theorem~\ref{thm:d1} with $\sigma=n$. We can assume without loss of generality that $\Phi$'s constraint circuits all have at least one input variable; otherwise it is a constant, so we evaluate all such circuits, output $0$ if any of them outputs $0$, and delete them otherwise. Then, since the density is constant, we have $\sum_j |S_j|=O(n)$, which implies that there are at most $O(n)$ constraint circuits $C_j$. 

From the circuit-CSP $\Phi$, we will build a bipartite graph with vertices on the left side corresponding to constraints and vertices on the right side corresponding to variables. More formally, for each constraint $(S_j,C_j)$ of $\Phi$ we add a vertex $v_j$ to our graph, and for each variable $x_i$ of $\Phi$, we add a vertex $u_i$. We add an edge between $v_j$ and $u_i$ iff $x_i\in S_j$. Set 
\[
  T_{v_j}(y)=C_j(y),
  \qquad
  T_{u_i}(y)=\begin{cases}1&\text{if all bits of $y$ are equal,}\\0&\text{otherwise.}\end{cases}
\]

We call each vertex $u_i$ with the tensor function $T_{u_i}$ defined above an \emph{equality vertex}, since it forces all assignments to its edges to be equal. This choice of $T_{u_i}$ means that whenever a term in the contraction is nonzero, an assignment to the edges corresponds exactly to an assignment of the original variables $x_i$ (since we do not allow $x_i$ to take on multiple values). Our choice of $T_{v_j}$ then further ensures that all circuits must evaluate to $1$ with this assignment. Thus, summing over all these assignments, the contraction $Z$ equals $\#\Phi$, the number of satisfying assignments to $\Phi$. (A variable that occurs in no constraint gives a vertex $u_i$ with no edges, which contributes a factor $1$ instead of $2$; we multiply $Z$ by $2$ for each such variable.) 

There is one issue with this construction: our assumption that the density is constant only bounds the \emph{average} degree of the $u_i$'s, but to apply~\Cref{thm:d1}, we need a bound on the \emph{maximum} degree of each vertex. We fix this by adding extra vertices. In the case that there is a $u_i$ with degree $d_i>n^{9/10}$, we replace $u_i$ with $d_i$ vertices $u_i^{(1)},\ldots,u_i^{(d_i)}$ and place an edge between $u_i^{(\ell)}$ and $u_i$'s $\ell$th neighbor for all $\ell\in[d_i]$. We then add $O(d_i)$ vertices to the other side of the bipartite graph and connect each of these vertices to two of the $u_i^{(\ell)}$'s such that there is a path in the graph between all the $u_i^{(\ell)}$'s. Then, if we make all these new vertices equality vertices, and use that equality is transitive, we can ensure that the assignments of the edges to all the $u_i^{(\ell)}$'s must be equal in order for that term in the contraction to be nonzero. 

This fixes the problem at the cost of $O(d_i)$ additional vertices and edges. Since the density is constant, we have $\sum_i d_i = \sum_j |S_j| = O(n)$. Therefore, in total, our graph will have $O(n)$ vertices and edges. Each $v_j$ has degree at most the arity $n^{9/10}$, each $u_i$ that was not split has degree at most $n^{9/10}$, and each new vertex has constant degree. Each equality vertex is trivial to compute, and each $T_{v_j}$ computes $C_j$, which takes $\poly(|\Phi|)$ time. Furthermore, the network is described by the graph and the circuits $C_j$, meaning $L=\poly(|\Phi|)$. Finally, $\#\Phi\le 2^n$, so computing the contraction modulo $2^{n+1}$ gives $\#\Phi$ exactly; we take $r=n+1$. Theorem~\ref{thm:d1} therefore computes $\#\Phi$ in time $2^{O(n/\sqrt{\log n})}\poly(|\Phi|)$.
\end{proof}

\section*{AI Disclosure}

OpenAI's Astra and Anthropic's Opus 5.5 and Fable 5.1 were used throughout our process.

\section*{Acknowledgments}

Bernardo is supported by NSF grant DMS-2434625. 
Carmen is supported by NSF awards 1942123, 2211237, 2224246, a Sloan
Research Fellowship, a Google Research Scholar Award, Omer Reingold’s Simons Foundation Investigator Award, Stanford Computer Science Distinguished
Fellowship, an NSF GRFP,  and a Jump
Trading PhD Fellowship. Any opinions, findings, and conclusions or recommendations expressed in
this material are those of the authors and do not reflect the views of Jump Trading.

 \bibliography{ref}
 \bibliographystyle{alphaurl}

\end{document}